\documentclass[english,aps,prapplied,twocolumn,superscriptaddress,longbibliography]{revtex4-2}
\usepackage[urlcolor=blue, hyperindex, colorlinks, bookmarks=true]{hyperref}

\usepackage{graphicx}
\usepackage{dcolumn}
\usepackage{bm}
\usepackage{amsmath,xcolor,amssymb}
\usepackage[normalem]{ulem}
\usepackage{bm}
\usepackage{soul}
\usepackage[utf8]{inputenc}
\usepackage[T1]{fontenc}
\usepackage{mathptmx}
\usepackage{mathtools}
\usepackage{float}
\RequirePackage[normalem]{ulem} 

\definecolor{new}{rgb}{.38,.6,.38}
\definecolor{old}{rgb}{1,0,0}
\definecolor{off}{rgb}{0,0,0}

\newcommand{\Buffalo}{\affiliation{Department of Physics, University at Buffalo, SUNY, Buffalo, NY~14260-1500, USA}}

\newcommand{\UCLA}{\affiliation{Department of Physics and Astronomy, University of California -- Los Angeles, Los Angeles, California 90095, USA}}
\newcommand{\CQSE}{\affiliation{Center for Quantum Science and Engineering, University of California -- Los Angeles, Los Angeles, California 90095, USA}}

\begin{document}
\title{Co-optimization of spin coherence and valley splitting in Si/SiGe heterostructures}

\author{Peihong Zhang}
\Buffalo
\author{Xuedong Hu}  
\Buffalo
\author{Saif Ullah}
\Buffalo
\author{Jason R. Petta}
\UCLA
\CQSE

\begin{abstract}
Single electron spins can be used to encode and process information in semiconductor quantum devices. Progress has been hindered by materials challenges, such as the small energy splitting between low-lying valley states and hyperfine coupling to nuclear spins. Here we use density functional theory to optimize the valley splitting and spin dephasing time in realistic Si/SiGe heterostructures.  Reductions in the Si quantum well width generally increase the valley splitting.  However, in narrow quantum wells, a larger fraction of the electronic wavefunction resides in the SiGe buffer layers, which increases the hyperfine coupling with spinful $^{73}$Ge.  Our work shows that Si/SiGe heterostructures with 3~--~4~nm wide quantum wells and $^{73}$Ge and $^{29}$Si concentrations of 50 ppm should support average valley splittings $E_{v}$~$>$~500~$\mu$eV and spin dephasing times $T_2^*$ exceeding 15~$\mu$s assuming an effective quantum dot area of 700 nm$^2$. In addition, sharper Si/SiGe interfaces in general result in larger valley splittings and longer spin dephasing times.  
\end{abstract}

\maketitle



All hardware platforms for quantum information processing have relative strengths and weaknesses. Superconducting qubits are straightforward to fabricate, but their large size may limit scaling and they suffer from short relaxation times \cite{Kjaergaard2020}. Trapped ions exhibit exceptionally long coherence times, but interaction energies are small and the resulting gate speeds are slow \cite{Bruzewicz2019}. While nitrogen vacancy centers in diamond support spin coherence at room temperature, the difficulty of engineering the precise placement of defect centers, combined with the challenge of optical control and readout, limits their applicability for quantum information processing \cite{childress_hanson_2013}.  

Semiconductor spin qubits encode quantum information in the electronic spin degree of freedom \cite{Burkard2023RMP}. Spin qubits are small ($\sim$100 nm) and can be lithographically fabricated using technologically mature materials such as silicon. Recent advances include demonstrations of high fidelity two-qubit gates \cite{Huang2019, mills_2Q_2021, Noiri2022, xue_2Q_2021}, scaling to larger quantum dot arrays \cite{Ha_archive, Neyens2024, Weinstein2022, undseth2026_weight_four, Abrahams2026_HRL}, as well as various approaches for microwave-based readout \cite{zheng_rapid_2019,Borjans_sensing_2021} and long-range spin-spin coupling \cite{Mi2018, Borjans2020, Samkharadze2018, Landig2018, Dijkema_NaturePhys}. To take full advantage of the scalability of semiconductor spin qubits, further improvements in coherence and reductions in device variability are required.

\begin{figure}[h!]
	\centering
	\includegraphics[width=\columnwidth]{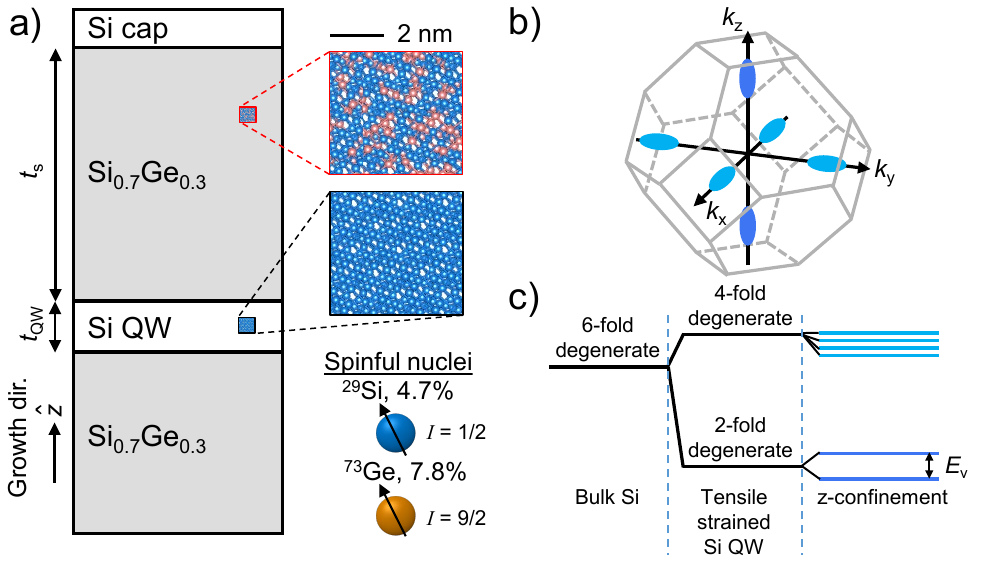}
	\caption{(a) Growth profile of a typical Si/SiGe heterostructure.  The silicon quantum well (QW) has a thickness $t_{\rm QW}$ and is buried by a Si$_{0.7}$Ge$_{0.3}$ spacer layer of thickness $t_{\rm s}$. Spinful $^{29}$Si and $^{73}$Ge nuclei result in electron spin decoherence through the contact hyperfine interaction. Inset: Atomic scale renderings of the Si and Si$_{0.7}$Ge$_{0.3}$ layers. (b) First Brillouin zone of bulk silicon with six equivalent valleys along the high-symmetry directions in $k$-space. (c) The six-fold valley degeneracy is lifted by tensile strain and confinement along the growth direction ($\hat{z})$.}
	\label{fig:hets}
\end{figure}

Of all of the semiconductor platforms being explored, silicon spin qubits are a leading contender for solid state quantum information processing due to its low natural abundance of spinful isotopes. While silicon provides a dramatic improvement over GaAs \cite{Petta2005}, where all of the lattice nuclei have nuclear spin $I$ = 3/2, the main source of spin decoherence in natural abundance Si quantum dots is still hyperfine coupling to the 5\%  concentration of $I$ = 1/2 $^{29}$Si \cite{Burkard2023RMP} [see Fig.~\ref{fig:hets}(a)].  Efforts to further improve spin dephasing times have thus focused on the isotopic enrichment of $^{28}$Si \cite{Muhonen2014,Eng2015}, and more recently on the isotopic depletion of $I$ = 9/2 $^{73}$Ge in the SiGe barriers \cite{Moutanabbir_AdvMat_2023}.

Low-lying valley states are a critical challenge facing Si spin qubits \cite{Burkard2023RMP}.  The bandstructure of bulk Si hosts six degenerate conduction band minima that are located close to the X point in the first Brillouin zone (FBZ), as shown in Fig.~\ref{fig:hets}(b).  In a strained Si/Si$_{0.7}$Ge$_{0.3}$ quantum well (QW) grown along the $\hat{z}$ direction, the $\pm z$ conduction band valleys are much lower in energy compared to the other four valleys \cite{schaffler1997}.  The sharp Si/SiGe interface potential, as well as alloy disorder in the barrier, couples the $\pm z$ valleys and lifts their degeneracy, giving rise to a valley splitting $E_{\rm v}$, as shown in Fig.~\ref{fig:hets}(c). A wide range of valley splittings have been reported in the literature for quantum dots in Si/Si$_{0.7}$Ge$_{0.3}$ QW, highlighting the critical role that interface abruptness and mesoscopic fluctuations have in defining this energy scale \cite{chen_detuning_2021}.  Small valley splittings make inter-valley transitions a threatening leakage channel for Si spin qubits, especially in tasks such as spin shuttling and exchange gates \cite{Volmer2024VSMap,Tariq_NPJQI2022}.  In other words, to ensure high fidelity spin qubit control, silicon quantum dots should exhibit consistently large valley splittings (e.g.~$E_{\rm v}$ $\gg$ $E_{\rm z}$, the Zeeman splitting, for Loss-DiVincenzo single spin qubits \cite{Loss1998}).

The perturbation potential that couples the $\pm z$-valleys giving rise to the valley splitting is primarily concentrated near the Si/SiGe interfaces \cite{boykin_valley_2004,friesen_theory_2010}. Most of the electronic wavefunction resides in the Si QW, with just the tails of the wavefunction penetrating into the SiGe barrier layers.  As a result, increasing the valley splitting is a challenging task. One strategy is to reduce the Si QW width, $t_{\rm QW}$, such as to increase the participation of the wavefunction in the SiGe barriers \cite{chen_detuning_2021}.  However, this approach increases the wavefunction overlap with $^{73}$Ge nuclei in the SiGe barriers. Alternative approaches to increasing the valley splitting include Ge spiking or precisely modulating the Ge concentration in the Si QW (i.e.~``wiggle well'') such as to maximally couple the $\pm z$ valleys \cite{mcjunkin2021valley,mcjunkinwigglewell}.  Again, both of these pathways increase coupling to $^{73}$Ge nuclei.  Therefore, the competing factors that influence the valley splitting and spin dephasing times must be co-optimized.

In this Letter, we present large-scale density functional theory (DFT) calculations of the valley states in  realistic Si/Si$_{0.7}$Ge$_{0.3}$ heterostructures, which allows us to evaluate both the valley splitting and the hyperfine coupling to residual nuclear spins in the Si QW as well as in the SiGe barriers.  Our objectives are to maximize the valley splitting by narrowing the Si QW and to clarify the requirements for isotopic enrichment of the Si QW and SiGe barriers such as to maximize the spin dephasing time $T_2^*$.  We find that reductions in the QW width on average increase $E_{\rm v}$. However, the increased participation of the wavefunction in the SiGe barriers increases the dephasing rate, which must be compensated for by isotopically depleting $^{73}$Ge and pursuing higher levels of $^{28}$Si isotopic enrichment to enable fault tolerant spin qubits with large valley splittings.

We use periodic SiGe/Si/SiGe superlattice structures to model Si QWs with varying width $t_{\rm QW}$ that is sandwiched between relaxed Si$_{0.7}$Ge$_{0.3}$ barriers. SiGe is an alloy and to accurately account for the effects of alloy disorder, we generate several large random structures containing thousands of atoms (i.e., $5\times 5\times 12$ cubic cells) using the special quasi-random structure (SQS) method \cite{schaffler1997,SQS1,SQS2}. For each random structure, we construct Si/SiGe heterostructures with $t_{\rm QW}$ = 2.15 -- 9.1 nm. The largest structures contain 5,800 atoms.  We then carry out DFT calculations using the Vienna Ab-initio Simulation Package (VASP) \cite{VASP1,VASP2}.

In total, we have studied more than 1,000 heterostructures with unique SiGe alloy disorder configurations, varying $t_{\rm QW}$, and the abruptness of the SiGe/Si interfaces. The PBEsol exchange-correlation energy functional is used for structural optimizations due to its accuracy in reproducing the structural parameters for solids \cite{PBEsol}. All structures are first optimized until the force on each atom is less than 0.005 eV/\AA. The optimized lattice constant for the Si$_{0.7}$Ge$_{0.3}$ random alloy is 5.507 \AA~(rescaled for an 8-atom cubic cell), which compares well with the experimental value of 5.497 \AA~for a fully relaxed alloy ~\cite{SiGe_Lattice1964}. For all subsequent calculations, the lateral lattice constants of the Si/SiGe structure are fixed at 5.507 \AA, while the perpendicular lattice constant (particularly in the Si QW) is allowed to relax.

The two nearly degenerate $\pm z$-valleys are the lowest energy conduction band states with wave functions confined within the QW. Accurately resolving small valley splittings ($\sim$ 10 $\mu$eV) requires a very high level of convergence in the calculations. The eigenvalues must converge to $\lesssim$ 1 $\mu$eV and the structures must also be fully relaxed. The calculated valley splitting can vary by more than 1~meV during the structural relaxation process. The $\lesssim$~1~$\mu$eV accuracy requirement for such large structures is computationally demanding.

\begin{figure}[t!]
	\centering
	\includegraphics[width=1\columnwidth]{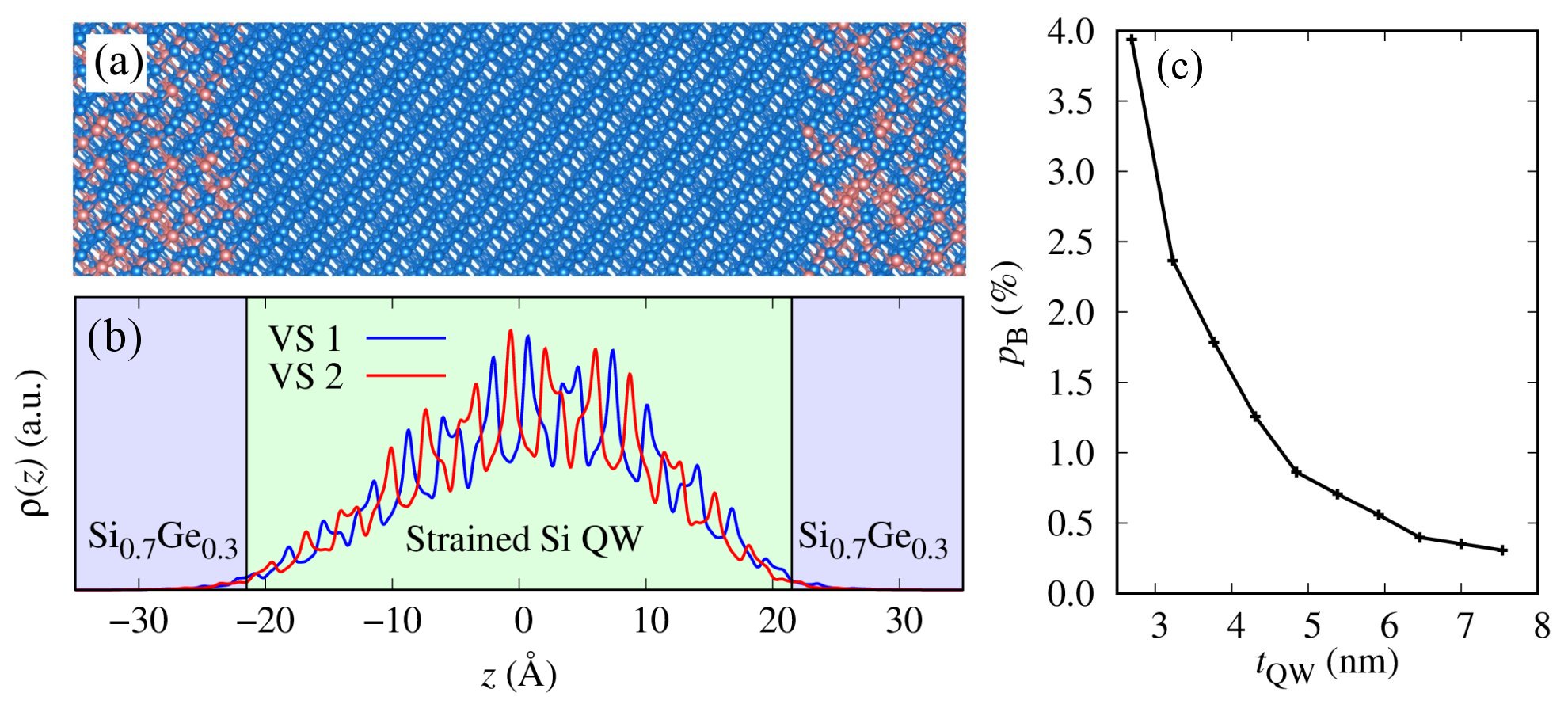}
	\caption{(a) An ideal Si/Si$_{0.7}$Ge$_{0.3}$ heterostructure with atomically abrupt interfaces.
(b) Charge density, $\rho(z)$, of the two valley states (VS 1 and VS 2) in the Si QW as obtained from DFT calculations using the heterostructure illustrated above. (c) Percentage of the electronic wavefunction residing in the SiGe barrier layers, $p_{\rm B}$, as a function of $t_{\rm QW}$. $p_{\rm B}$ rapidly increases as $t_{\rm QW}$ is decreased below $\sim$~5~nm, exacerbating the effects of hyperfine coupling to spinful $^{73}$Ge nuclei.
	}
	\label{fig:Wavefunctions}
\end{figure}

Contact hyperfine coupling calculations require knowledge of the electronic wavefunction at the nuclear sites, which is not accessible with the usual pseudopotential method \cite{Taylor2007}. Conventionally, all-electron calculations are required. Here we use the projector augmented wave (PAW) method \cite{PAW_PhysRevB.50.17953, PAW-hyperfine-PhysRevB.62.6158}, which allows the reconstruction of the all-electron spin density as implemented in VASP \cite{VASP1,VASP2}.

Figure~\ref{fig:Wavefunctions}(a) shows an example of the atomic structure of a Si QW with $t_{\rm QW}$ = 4.3 nm  sandwiched between Si$_{0.7}$Ge$_{0.3}$ random alloy barriers with sharp Si/SiGe interfaces.  The resulting charge density ${\rho}(z)~=~\int|\psi(\textbf{r})|^2dxdy$ of the two lowest energy valley states is plotted in Fig.~\ref{fig:Wavefunctions}(b).  Qualitatively, both the valley splitting and the electronic hyperfine coupling with the nuclear spins in the SiGe barriers is sensitive to the tails of the electronic wavefunction extending into the SiGe barriers.  In Fig.~\ref{fig:Wavefunctions}(c), we plot the percentage of the ground state electronic wavefunction residing in the Si$_{0.7}$Ge$_{0.3}$ barriers, $p_{\rm B}$, as a function of $t_{\rm QW}$. As $t_{\rm QW}$ decreases below $\sim$ 5 nm, $p_{\rm B}$ rapidly increases, as does the hyperfine coupling to $^{73}$Ge nuclear spins.

\begin{figure}[t!]
	\centering
	\includegraphics[width=\columnwidth]{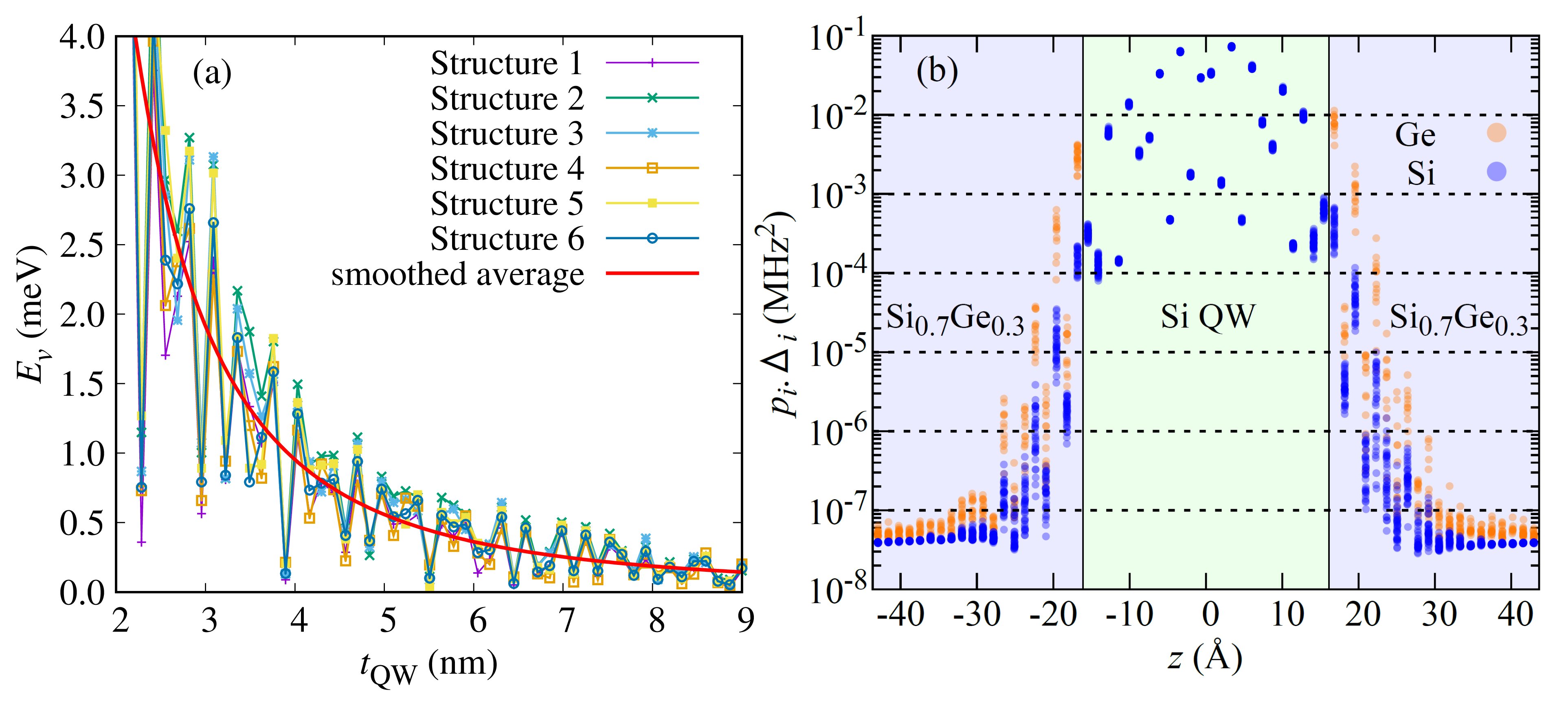}
	\caption{(a) $E_{\rm v}$ as a function of $t_{\rm{QW}}$ for idealized structures with atomically abrupt interfaces.  The solid red line depicts the average valley splitting $\langle E_{\rm v} \rangle$. (b) Contributions from individual nuclear spins to the contact hyperfine coupling in a QW with $t_{\rm{QW}}$ = 3.2 nm. The blue (orange) dots represent contributions from $^{29}$Si ($^{73}$Ge) lattice nuclei.}
	\label{fig:ValleySplittings}
\end{figure}

In Fig.~\ref{fig:ValleySplittings}(a), we plot the valley splitting $E_{\rm v}$ as a function of $t_{\rm{QW}}$ for six different heterostructures with ideal (atomically abrupt) interfaces. The increment of $t_{\rm{QW}}$ is one Si atomic layer (1.34 \AA, i.e., 1/4 of the $z$ lattice constant of the strained Si. Our simulations show that it is possible to achieve $E_{\rm v}$ $>$ 1 meV for $t_{\rm QW}$ $<$ 4 nm.  On the other hand, for wider QWs (e.g.,~$t_{\rm{QW}}$ $\geq 8$ nm), $E_{\rm v}$ $\approx$ 0.1 -- 0.25 meV.  In addition to the overall decrease in $E_{\rm v}$ with increasing $t_{\rm{QW}}$, the valley splitting exhibits large fluctuations with small changes in $t_{\rm{QW}}$. Our findings are consistent with experimental reports of the valley splitting varying over a wide range in multi-quantum dot devices \cite{Mills2019, Volmer2024VSMap}. For $t_{\rm{QW}}$ = 6 -- 9 nm, even in these ideal structures with abrupt interfaces, valley splittings can be as small as 30 $\mu$eV.  To better illustrate the general trend of $E_{\rm v}$ with $t_{\rm{QW}}$, we average the calculated valley splitting over half of the silicon lattice constant for each of the Si/SiGe heterostructures.  The resulting average, $\langle E_{\rm v} \rangle$, is plotted as the red solid curve in Fig.~\ref{fig:ValleySplittings}. There is a clear trend of $\langle E_{\rm v} \rangle$ increasing with decreasing $t_{\rm{QW}}$.

The tails of the electronic wavefunction penetrating into the SiGe barriers [see Figs.~\ref{fig:Wavefunctions}(b,c)] result in additional contributions to the contact hyperfine interaction with $^{29}$Si and $^{73}$Ge nuclear spins.  The hyperfine interaction between an electron and a nuclear spin at position $\mathbf{r}$ is dominated by the Fermi contact term,
\begin{equation} 
A = \frac{4}{3} \mu_0 g_e\mu_B g_N\mu_N|\psi(\mathbf{r})|^2,
\end{equation}
where $\mu_0$ is permeability of free space, $\mu_B$ and $\mu_N$ are the electron and nuclear magnetons, $g_e$ and $g_N$ are the electron and nuclear $g$-factors, and $\psi(\mathbf{r})$ is the electronic wavefunction \cite{Taylor2007}. The small nuclear magneton ensures that at typical magnetic fields ($B$~<~2~T), the nuclear spins are in a thermal state and randomly oriented. Through the hyperfine interaction, their randomness results in an energy spread of the electron spin state
\begin{equation} 
\langle \delta E^2 \rangle = \sum_i p_i \frac{I_i(I_i+1)}{3}|A_i|^2 = \sum_i p_i \Delta_i,  
\label{eq:HFEnergySpread}
\end{equation}
where the summation runs over all nuclear sites $i$, with $p_i$ being the abundance of a spinful isotope, and $I_i$ its spin.  In Fig.~\ref{fig:ValleySplittings}(b) we plot $p_i\Delta_i$ for a Si/SiGe heterostructure with natural isotope abundance, where each point represents a contribution to the hyperfine coupling from an individual spinful nucleus.  At an equivalent location (in terms of wavefunction magnitude), the Ge contributions are nearly two orders of magnitude larger than those of Si since $I = 9/2$ for $^{73}$Ge.  With $p_i\Delta_i$ being plotted on a log-linear graph, it is clear that the overall hyperfine coupling is dominated by nuclei inside the QW even for a narrow well with $t_{\rm QW}$ = 3.2 nm. However, with high levels of $^{28}$Si enrichment in the QW, we can anticipate that contributions from $^{73}$Ge in the SiGe barriers will become significant. We next examine the dependence of the spin dephasing time $T_2^*$ on the level of Ge and Si isotopic enrichment. 

Fluctuations in the electron Zeeman splitting caused by hyperfine coupling to spinful lattice nuclei (see Eq.~\ref{eq:HFEnergySpread}) lead to electron spin dephasing, which is described by the inhomogeneous spin dephasing time \cite{Petta2005}:
\begin{equation} 
T_2^\star=\hbar \sqrt{\frac{2}{\langle \delta E^2\rangle}} \sqrt{\frac{A_{\rm{QD}}}{A_{\rm{model}}}}. 
\end{equation} 
Note that in our calculations the electronic wavefunction is normalized within the volume of the model ($\leq$ 5800 atoms). The area of the model structure $A_{\rm model}$ = 7.6 nm$^2$ 
is limited to make the DFT calculations computationally tractable.  Our results must therefore be rescaled to match the lateral effective area of the quantum dot in experiment. Here we assume a small 
effective area $A_{\rm QD}$ = 700 nm$^2$, which is obtained through electrostatic modeling of gate-defined Si/SiGe QDs \cite{Anderson_Gyure,Zajac2016}.   

\begin{figure*}[th]
	\centering
	\includegraphics[width=1.01\textwidth]{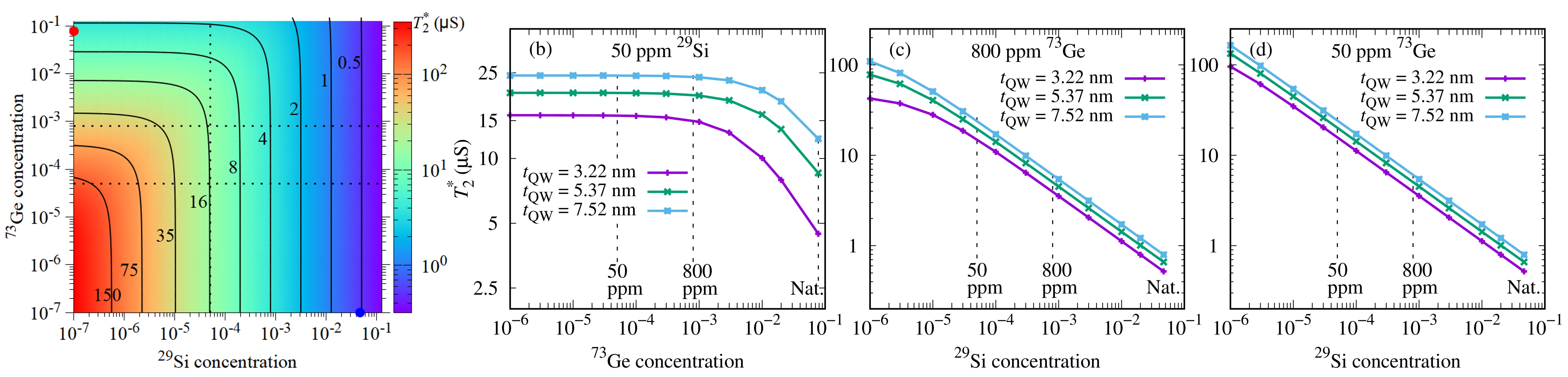}
	\caption{(a) Color-scale plot of the inhomogeneous spin dephasing time $T_2^*$ as a function of $^{29}$Si and $^{73}$Ge concentrations. (b--d) Linecuts extracted from panel (a). In panel (b), $T_2^*$ is plotted as a function of $^{73}$Ge concentration for 50 ppm $^{29}$Si.  In panels (c,d), $T_2^*$ is plotted as a function of $^{29}$Si concentration for 800 ppm and 50 ppm $^{73}$Ge.}
	\label{fig:Dephasing}
\end{figure*}

In Fig.~\ref{fig:Dephasing}(a), $T_2^{\star}$ is plotted as a function of both $^{29}$Si concentration (in the QW and SiGe barriers) and $^{73}$Ge concentration (in the SiGe barriers) for a structure with $t_{\rm QW}$ = 3.22 nm.  It is notable that when the  $^{29}$Si concentration exceeds $\sim$1000 ppm, $T_2^{\star}$ has almost no dependence on the $^{73}$Ge concentration in the SiGe barriers, as the $^{29}$Si nuclear spins in the QW are the dominant contributors to electron spin dephasing.  When the $^{29}$Si concentration is reduced to a much lower level, $^{73}$Ge nuclei in the SiGe barriers gradually become a significant contributor to spin dephasing.  

Figures~\ref{fig:Dephasing}(b--d) summarize the dependence of $T_2^*$ on isotopic purity for structures with $t_{\rm QW}$ = 3.22, 5.37, and 7.52 nm. In Fig.~\ref{fig:Dephasing}(b), we plot $T_2^{\star}$ as a function of the $^{73}$Ge concentration in the SiGe barriers with a $^{29}$Si concentration of 50 ppm (the highest level of isotopic enrichment commonly available for large scale production of Si/SiGe wafers). With 50~ppm of $^{29}$Si, the isotopic concentration of $^{73}$Ge in the SiGe barriers is now important: as it is reduced from natural abundance ($\sim$ 7.7\%) down to 0.1\%, $T_2^{\star}$ increases for all three QW widths.  For example, with $t_{\rm QW}$ = 3.22 nm, $T_2^{\star}$ triples, from roughly 4 $\mu$s up to 13 $\mu$s. For $^{73}$Ge concentrations less than 10$^{-4}$, further isotopic depletion of $^{73}$Ge is not effective as spin dephasing is again dominated by $^{29}$Si in the QW. 

Figures~\ref{fig:Dephasing}(c,d) illustrate the impact of silicon purification with 800 ppm and 50 ppm of $^{73}$Ge in the barriers, respectively.  With the effects of the barrier $^{73}$Ge nuclear spins suppressed, $T_2^*$ now monotonically increases as the concentration of $^{29}$Si is reduced. However, for $t_{\rm QW}$ = 3.22 nm and 800 ppm of $^{73}$Ge in the SiGe barriers, $T_2^{\star}$ begins to saturate when the $^{29}$Si concentration is reduced to $\leq$10 ppm, demonstrating that the dephasing effect of $^{73}$Ge becomes important again.  In short, Fig.~\ref{fig:Dephasing} unambiguously shows that to enhance $T_2^{\star}$, the isotopic purification of $^{73}$Ge in the barriers must go hand in hand with further enrichment of $^{28}$Si. Such purification is particularly important for thinner quantum wells.

The results presented above are for Si/SiGe heterostructures with atomically sharp interfaces. However, Si/SiGe interfaces produced by chemical vapor deposition and molecular beam epitaxy are not atomically abrupt due to the significant lattice constant mismatch between Si and Si$_{0.7}$Ge$_{0.3}$ \cite{Deelman_MRS,schaffler1997}. To model the effects of interfaces with finite widths, we include a 1.1 nm thick Si$_{0.85}$Ge$_{0.15}$ transition layer on both sides of the Si QW, and then re-calculate the electronic structure of the heterostructures.

\begin{figure}[b!]
	\centering
	\includegraphics[width=\columnwidth]{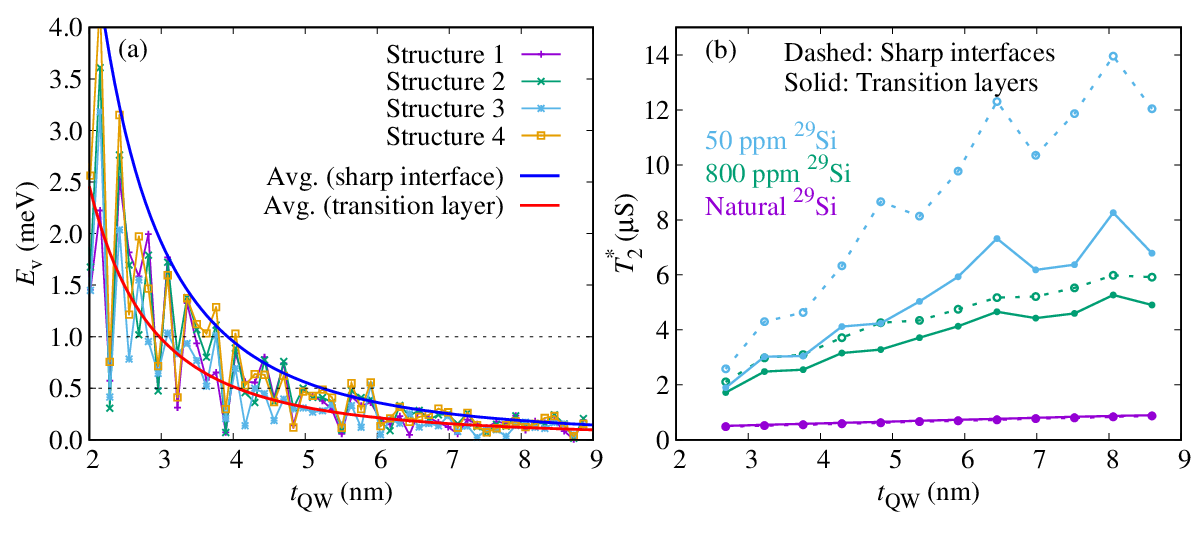}
	\caption{(a) $E_{\rm v}$ for QWs bounded by 1 nm thick Si$_{0.85}$Ge$_{0.15}$ transition layers.  Results from four different SiGe structures are shown.  The red solid line shows the average valley splitting $\langle E_{\rm v} \rangle$ for these structures. For comparison, the results from Fig.~(\ref{fig:ValleySplittings}) are plotted here as well (blue line, abrupt interface). (b) A comparison of $T_2^\star$ as a function of $t_{\rm QW}$ for abrupt interfaces (dashed lines) and broadened interfaces (solid lines).}
	\label{fig:TransitionLayer}
\end{figure}

The resulting valley splittings and spin dephasing times are plotted in Fig.~\ref{fig:TransitionLayer}. Comparing with Fig.~\ref{fig:ValleySplittings}(a), the transition layer structures exhibit the same alloy-structure-dependent randomness in $E_{v}$ and significant fluctuations with $t_{\rm QW}$, as well as the clear trend of increasing $E_{v}$ with decreasing $t_{\rm QW}$.  Numerically, however, the insertion of the transition layer leads to a nearly 50\% reduction in $E_{v}$ for all of the $t_{\rm QW}$ investigated here.

Figure~\ref{fig:TransitionLayer}(b) shows $T_2^*$ as a function of $t_{\rm QW}$ for atomically abrupt interfaces (dashed lines) and interfaces that include the transition layer (solid lines).   For simplicity, the $^{73}$Ge in the SiGe barriers is 7.7\% (natural abundance). If the Si is also at natural abundance (4.7\%), as shown by the solid purple curve, there is a slow rise of $T_2^*$ with increasing $t_{\rm QW}$ due to the $1/\sqrt{N}$ dependence of $T^*_2$ on the number $N$ of spinful nuclei within the quantum dot \cite{Taylor2007}.  In this configuration, the interface feature has a vanishingly small effect: spin dephasing is completely dominated by the $^{29}$Si nuclear spins in the QW.  On the other hand, when the $^{29}$Si concentration is reduced down to 50~ppm, the inclusion of the transition layer has a notable effect.  The softened confinement potential gives rise to a larger contribution of $^{73}$Ge spins in the barrier and thus shortens
$T_2^*$.  The inclusion of the transition layer also leads to a more rapid decrease in $T_2^*$ as $t_{\rm QW}$ decreases.  In summary, the results from our simple model show that a broadened interface is always worse than a sharp interface in the context of electron spin properties -- both the valley splitting and dephasing time are reduced.  As such, the refinement of Si/SiGe growth processes can yield dividends in two important quantum dot performance metrics - $T_2^*$ and $E_v$ \cite{scappucci_germanium_2020}.

In conclusion, our DFT calculations provide guidance for the levels of isotopic enrichment that are required to support inhomogeneous spin dephasing times $T_2^*$~$>$~20~$\mu$s and consistently large valley splittings $E_v$~$>$~200~$\mu$eV in Si/SiGe heterostructures. In contrast with past studies \cite{CvitLoss}, we carry out large-scale first-principles calculations of realistic Si/SiGe heterostructures with varying QW widths and include narrow ($\sim$3~nm) quantum well models where the valley splitting is expected to be the largest. We also consider transition SiGe layers to more accurately simulate heterostructures with imperfect Si/SiGe interface abruptness.  Broadening of the Si/SiGe interface regions has a deleterious impact on both $E_{\rm v}$ and $T_2^*$.  Therefore, improvements in the growth of Si/SiGe heterostructures can positively benefit two important figures of merit for silicon spin qubits.

\begin{acknowledgements}
We thank useful discussion with M. Gyure, C. Anderson, 
V. Lordi, and J. B. Varley. We acknowledge the support of AFOSR grant FA9550-23-1-0710 and ARO grants W911NF-23-1-0104 and W911NF-23-1-0018.

\end{acknowledgements}

\newpage

\bibliography{RMP_master2_bib_v4}

\end{document}